\documentclass[reprint,twocolumn,superscriptaddress,nofootinbib,amsmath,amssymb]{aastex63}
\pdfoutput=1
\usepackage{graphicx}
\usepackage{bm}
\usepackage{amsmath}
\usepackage{amssymb}
\usepackage{color}
\usepackage{colortbl}
\usepackage{hyperref}
\usepackage{times}

\newcommand{\cf}{cf.,~}
\newcommand{\ie}{i.e.,~}
\newcommand{\eg}{e.g.,~}

\definecolor{orange}{rgb}{1.0, 0.5, 0.0}
\definecolor{gray}{gray}{0.9}

\newcommand{\Mth}{M_{\rm th}}

\graphicspath{{figs/}}

\begin{document}

\title[Quasi-universal behaviour of the threshold mass]{Quasi-universal
  behaviour of the threshold mass in unequal-mass, spinning binary
  neutron-star mergers}

\author[0000-0001-9781-0496]{Samuel D. Tootle}
\affiliation{Institut f\"ur Theoretische Physik, Goethe Universit\"at, 
Max-von-Laue-Str. 1, 60438 Frankfurt am Main, Germany}
\email{tootle@itp.uni-frankfurt.de}
\author[0000-0002-6400-2553]{L. Jens Papenfort}
\affiliation{Institut f\"ur Theoretische Physik, Goethe Universit\"at,
Max-von-Laue-Str. 1, 60438 Frankfurt am Main, Germany}
\author[0000-0002-0491-1210]{Elias R. Most}
\affiliation{Princeton Center for Theoretical Science, Princeton University, Princeton, NJ 08544, USA}
\affiliation{Princeton Gravity Initiative, Princeton University, Princeton, NJ 08544, USA}
\affiliation{School of Natural Sciences, Institute for Advanced Study, Princeton, NJ 08540, USA}
\author[0000-0002-1330-7103]{Luciano Rezzolla}
\affiliation{Institut f\"ur Theoretische Physik, Goethe Universit\"at,
Max-von-Laue-Str. 1, 60438 Frankfurt am Main, Germany}
\affiliation{School of Mathematics, Trinity College, Dublin 2, Ireland}
\affiliation{Frankfurt Institute for Advanced Studies, Ruth-Moufang-Str. 1, 60438 Frankfurt am Main, Germany}

\begin{abstract}
The lifetime of the remnant produced by the merger of two neutron stars
can provide a wealth of information on the equation of state of nuclear
matter and on the processes leading to the electromagnetic
counterpart. Hence, it is essential to determine when this lifetime is
the shortest, corresponding to when the remnant has a mass equal to the
threshold mass, $M_{\rm th}$, to prompt collapse to a black hole. We
report on the results of more than 360 simulations of merging
neutron-star binaries covering 40 different configurations differing in
mass ratio and spin of the primary. Using this data, we have derived a
quasi-universal relation for $M_{\rm th}$ and expressed its dependence on
the mass ratio and spin of the binary. The new expression recovers the
results of \citet{Koeppel2019} for equal-mass, irrotational binaries and
reveals that $M_{\rm th}$ can increase (decrease) by $5\%~(10\%)$ for
binaries that have spins aligned (anti-aligned) with the orbital angular
momentum and provides evidence for a non-monotonic dependence of $M_{\rm
  th}$ on the mass asymmetry in the system. Finally, we extend to
unequal-masses and spinning binaries the lower limits that can be set on
the stellar radii once a neutron-star binary is detected, illustrating
how the merger of an unequal-mass, rapidly spinning binary can
significantly constrain the allowed values of the stellar radii.
\end{abstract}


\section{Introduction}
\label{sec:intro}

With the birth of multi-messenger gravitational-wave astronomy and the
detection of binary neutron star (BNS) mergers via the the GW170817 event
\citet{Abbott2017}, and the use of numerical simulations, it has become
possible to establish a number of constraints on the equation of state
(EOS) of nuclear matter \citep[see,
  \eg][]{Margalit2017,Bauswein2017b,Shibata2017c,Rezzolla2017,Ruiz2017,Radice2017b,Montana2018,Raithel2018,Tews2018,Malik2018}. Due
to the degeneracy between tidal and stellar spin effects on the inspiral
waveform, low- or high-spin priors are needed to infer the properties of
the binaries from the gravitational-wave measurements
\citep{Abbott2017}. Strong emphasis has been given to the low-spin priors
under the expectation that the neutron stars (NSs) lose a significant
fraction of their spin angular momentum through electromagnetic dipolar
radiation well before the merger. While these assumptions favor
irrotational NSs in the late inspiral and merger of BNSs, the parameter
estimations from the gravitational-wave detections still strongly depend
on the given \textit{a-priori} distribution of expected spins, with the
high-spin prior leading to large uncertainties in the mass ratio and
effective spin of the binary \citep{Abbott2020}.

Much of the theoretical modelling of BNS mergers has been concentrated on
mass ratios $q:=M_2/M_1\gtrsim0.7$ and on irrotational constituents
\citep[see,
  \eg]{Shibata06a,Rezzolla:2010,Bauswein2013,Dietrich:2015b,Radice2016,Lehner2016},
in accordance with the (limited) sample of observed binary pulsar
systems. However, the modelling of smaller ratios $q\gtrsim0.45$
\citep{Dietrich2017,Most2020e,Papenfort2021b} and of higher spins has
also started
\citep{Kastaun2013,Bernuzzi2013,East2016,Dietrich2017c,Most2019}. Given
its impact on the electromagnetic counterpart, a particularly important
prediction of the theoretical modelling has been the determination of
whether the BNS underwent a prompt collapse at merger. The threshold mass
discerning a prompt from a delayed collapse has been investigated
thoroughly for irrotational binaries \citep[see][for some
  reviews]{Baiotti2016,Burns2020} using a number of EOSs. The natural
expectation that $\Mth$ can be parametrized in terms of the
maximum mass of a nonrotating NS, $M_{_{\rm TOV}}$ \citep{Bauswein2013},
has been refined by more advanced parametrizations
\citep{Koeppel2019,Agathos2019} and by incorporating the effect of
asymmetric binary systems \citep{Bauswein2020c}. The latter resulted in a
lowering of $\Mth$ at small mass ratios, \ie $q\gtrsim0.6$,
depending on the stiffness of the EOS \citep{Bauswein2020c}. Furthermore,
simulations have suggested that the lifetime of merger remnants increases
with the binary spin
\citep{Kastaun2013,Bernuzzi2013,East2016,Kiuchi2019}, and decreases
significantly at very small mass ratios
\citep{Rezzolla:2010,Dietrich2017c,Kiuchi2019,Bernuzzi2020}.

Using three temperature-dependent EOSs compatible with present
astronomical observations, we report on a systematic exploration the
effects that spin and mass asymmetry have on $\Mth$ of BNS
configurations with mass ratio in the range $ 0.5 \leq q \leq 1$,
concentrating mostly on configurations in which the primary is spinning,
while the secondary is irrotational. The large amount of data collected in
this way has allowed us to derive a quasi-universal relation for $M_{\rm
  th}$ in terms of the mass ratio and spin of the binary.

\section{Numerical Methods and Initial Data}
\label{sec:nummethods}

\noindent\textit{Numerical Methods.~} A significant hurdle to exploring
asymmetric spinning binaries is the construction of initial data (ID)
consistent with the Einstein equations. Here, we have used \texttt{FUKA}
\citep{Papenfort2021b}, the first publicly available ID solver able to
reliably explore the needed parameter space. See the Appendix
  Materials (AM) for details on the configurations considered. The
evolution of the ID is performed using the \texttt{Einstein Toolkit}
infrastructure which includes a fixed-mesh box-in-box refinement
framework \texttt{Carpet} \citet{Schnetter-etal-03b}. The spacetime
evolution was handled by \texttt{Antelope} \citep{Most2019b} using a
constraint damping formulation of the Z4 system
\citep{Bernuzzi:2009ex,Alic:2011a}.  Finally, the general-relativistic
magnetohydrodynamic code, \texttt{FIL} \citep{Most2019b}, was used to
evolve the fluid quantities. \texttt{FIL} is a derivative of the original
\texttt{IllinoisGRMHD} code \citep{Etienne2015} with the addition of
high-order (fourth) conservative finite-differencing methods
\citep{DelZanna2007}. Importantly, \texttt{FIL} handles EOSs that are
dependent on temperature and electron-fraction, and includes a neutrino
leakage scheme to handle neutrino cooling and weak interactions. The
simulations were performed using a grid setup with an extent of
$\approx3000\,{\rm km}$, decomposed in six levels of refinement, with the
finest grid spacing being $\Delta x\approx295\,{\rm m}$. For comparison,
36 simulations of higher-resolution were performed with a finer grid
spacing of $\Delta x\approx221\,{\rm m}$, resulting in a measure of the
error budget of $\sim1\%$ ($\Delta\Mth\simeq0.03\,M_\odot$).

\begin{figure*}
  \centering
  \includegraphics[width=0.32\textwidth]{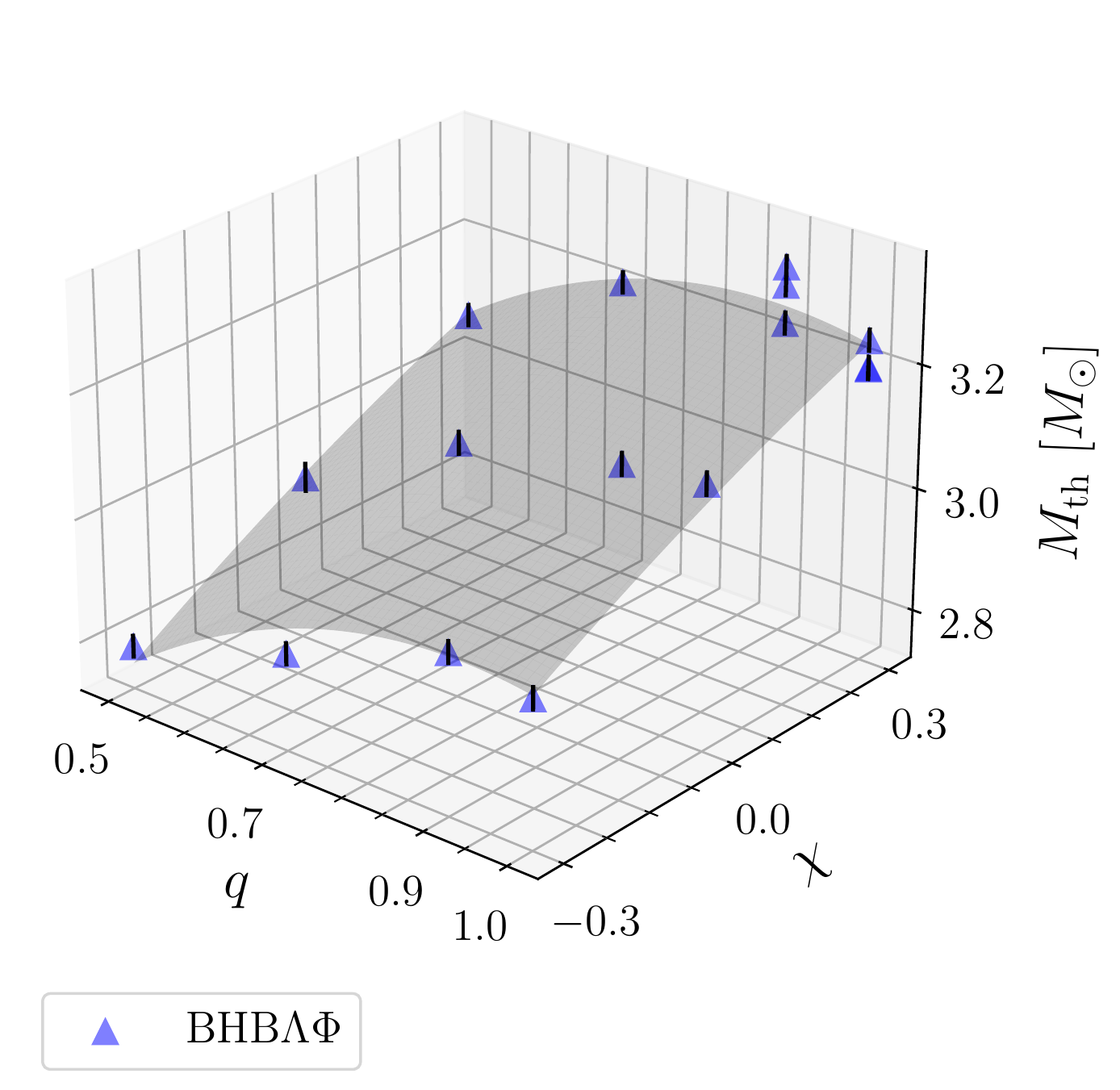}
  \includegraphics[width=0.32\textwidth]{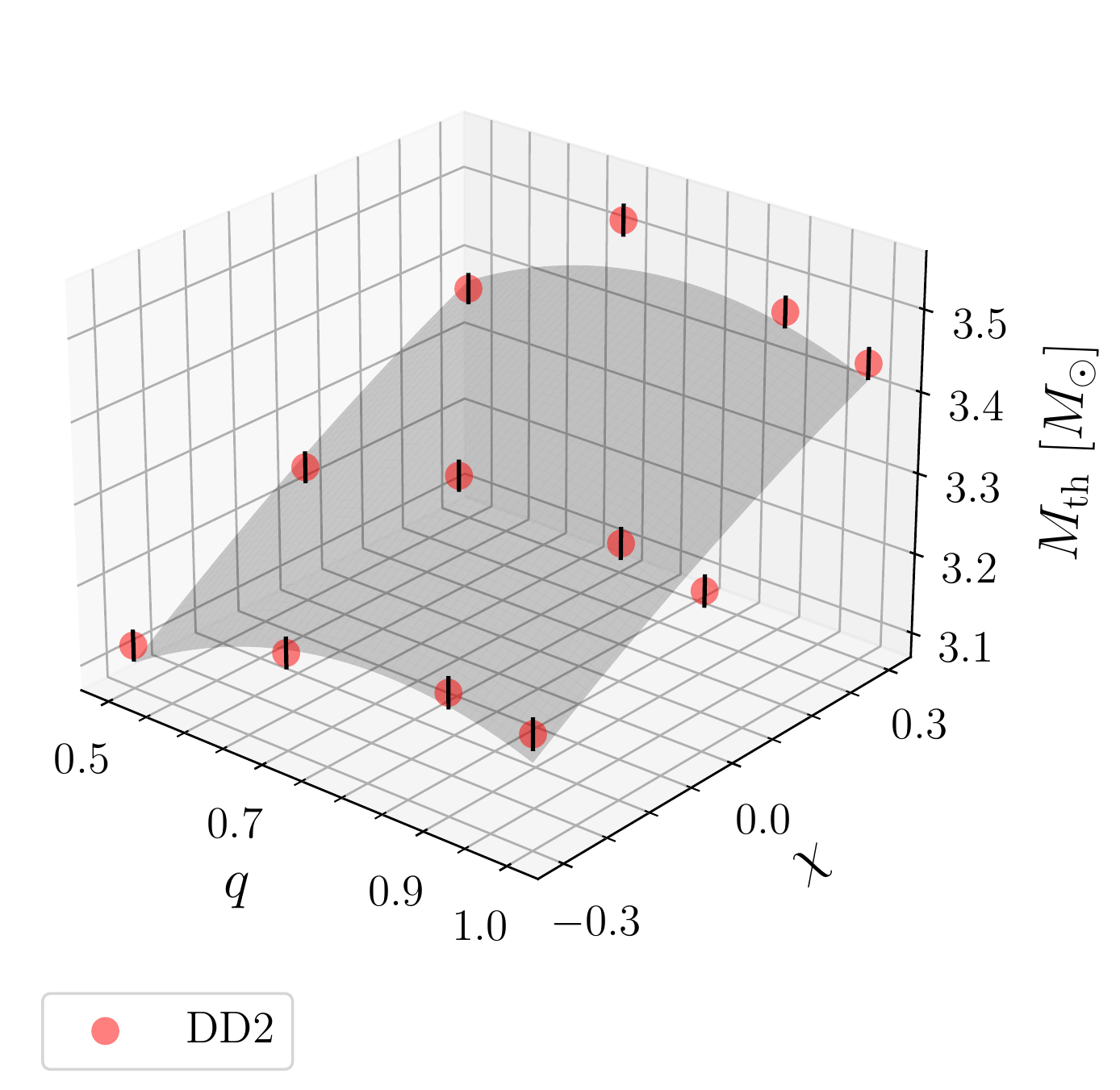}
  \includegraphics[width=0.32\textwidth]{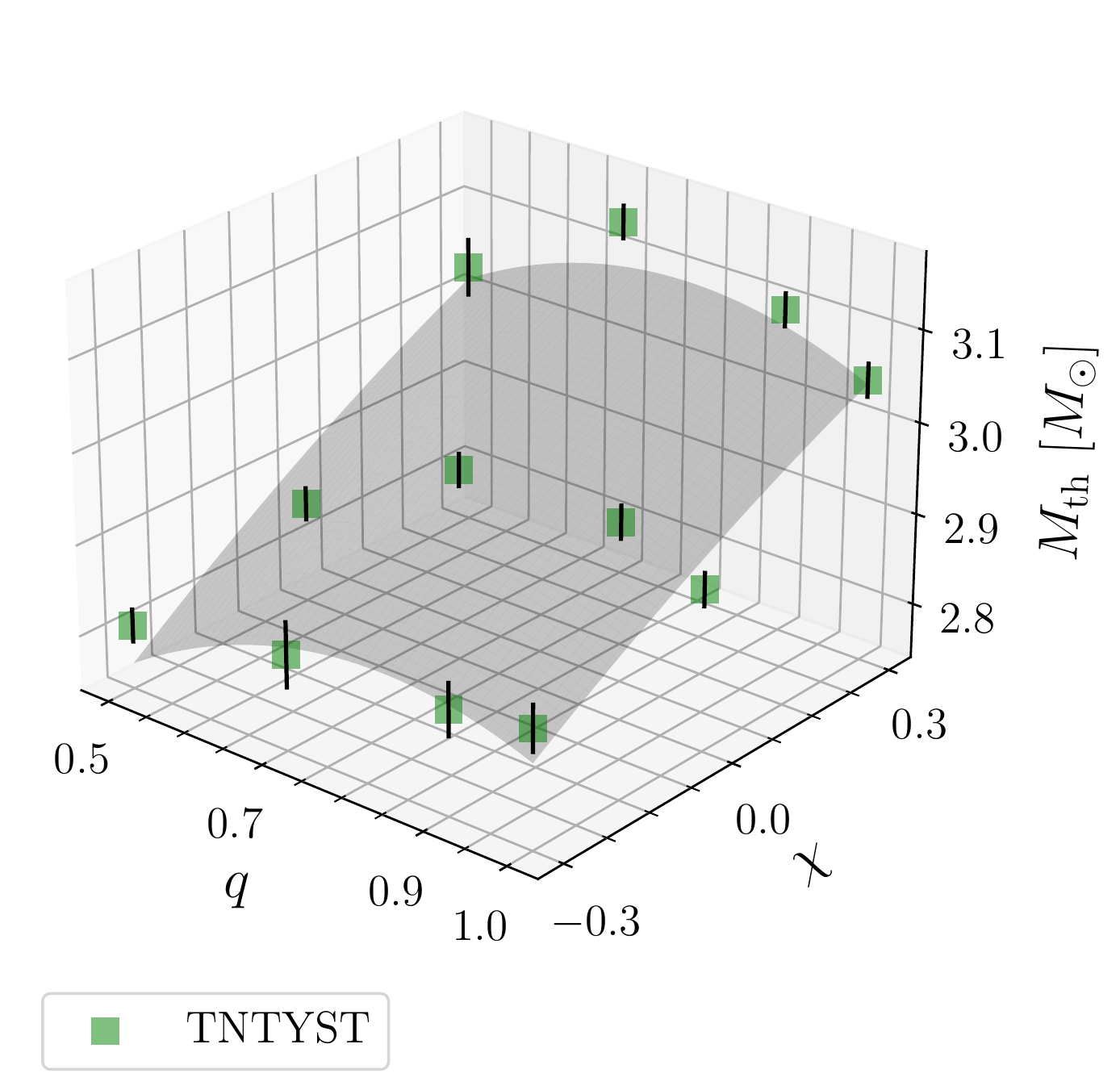}
  \caption{Dependence of $\Mth$ on the mass ratio and total
    spin of the binary. The three panels depict the data for the
    EOSs considered here, while the gray-shaded surface represents
    the fit via the quadratic ansatz \eqref{eq:f_ansatz}.}
  \label{fig:3dfit}
\end{figure*}

\noindent\textit{Space of parameters.~} Pulsar observations have provided
a wealth of information on the various rotation states that NSs can have
including isolated pulsars with extreme spins and
binary configurations with moderate to high spins. In general, most of
the spin angular momentum of the binary is associated with the most
massive of the two NSs, \ie the primary \citep{Lazarus2016}. Hence, the
spin configurations considered here are mostly centered on the effects 
$\chi_1$ has on $M_{\rm th}$. Here, 36 out of the 40
binaries considered have $\chi_1=[-0.3, 0, 0.3]$ and $\chi_2=0$
 where $\chi_{1,2}:=J_{1,2}/M^2_{1,2}$ are, respectively, the dimensionless
spins of the primary and secondary with spin angular momenta
$J_{1,2}$. However, in order to gauge the impact of the spin of the
secondary we have considered also four binary configurations with the
same effective dimensionless spin
$\tilde{\chi}:=(1+q)^{-1}\,\left(\chi_1+q\,\chi_2\right)$.

\noindent\textit{Collapse-Time Measurement.~} A crucial aspect of any
study wishing to determine $M_{\rm th}$ is a rigorous, systematic and
reproducible definition of what constitutes a collapse.  Such an approach
was previously not always considered, and very qualitative definitions of
the threshold mass have been employed in the literature. Here,
  instead, we follow the prescription proposed by \citet{Koeppel2019},
  which tracks the minimum of the lapse to compute the collapse time and
  compares it to the shortest possible over which a prompt collapse can
  take place, that is, the free-fall timescale. This approach, whose
details can be found in the AM, allows not only for a precise definition
and measurements, but also for the reproducibility of the results
presented here.

\section{Results}
\label{sec:results}

\noindent\textit{Dependence on Mass Ratio and Spin.~} While the work of
\citet{Koeppel2019} provided a first well-defined and rigorous manner of
determining a quasi-universal relation for $M_{\rm th}$, it was
restricted to the analysis of equal mass, \ie with $q=1$ and irrotational
binaries, \ie with $\chi_1=0=\chi_2$. In this specific scenario, $M_{\rm
  th}$ was found to depend on the EOS rather simply, so that a
quasi-universal relation $\Mth=\Mth({\rm EOS})$ was proposed. However, it
is clear that when considering the additional influence of spin and mass
asymmetry, the functional dependence of $M_{\rm th}$ must account also
for these additional degrees of freedom so that $\Mth=\Mth({\rm
  EOS},q,\chi)$, where $\chi:=\chi_1+\chi_2$ is the total dimensionless spin
of the binary.

Determining the {\it exact} expression for $\Mth({\rm EOS},q,\chi)$
clearly requires the exploration of the space of parameters in mass ratio
and spin for any given EOS. As an example, we report in
Fig. \ref{fig:3dfit} the dependence on the total spin and mass ratio of
$M_{\rm th}$ for the three EOSs considered here. More specifically, shown
with different coloured symbols are the values of $M_{\rm th}$ -- and the
corresponding error bars, which are shown in black -- as determined
following the prescription discussed in Sec. \ref{sec:nummethods} and by
\citet{Koeppel2019} (see also the AM for an alternative approach leading
to the same results), while the gray-shaded surfaces represent the best
fitting function to the data. Note that the fits for the three EOSs have
variable errors, but all with a small chi-squared of
$X^2_{\texttt{BHB}\Lambda\Phi}=0.001, X^2_{\texttt{DD2}}=0.001$, and
$X^2_{\texttt{TNTYST}}=0.003$, all of which have an
  average (maximum) deviation from the fit that is $1\%~(2\%)$. 
  When expressed in absolute terms, the average deviations from the 
  fits amount to $\Delta M_{{\rm th}}=0.03\,M_{\odot}$.

\noindent\textit{Quasi-Universal Behaviour.~} As can be readily
appreciated from the inspection of Fig. \ref{fig:3dfit} -- which
correspond to 40 distinct binary configurations differing in mass ratio
and spin -- $M_{\rm th}$ shows a behaviour that is similar for the three
EOSs, but also that it leads to systematically different values for each
of the EOSs considered. Determining accurately the threshold mass for
each configuration has required the calculation of the inspiral and
merger of about a dozen simulations with varying initial mass, the
computational cost associated with Fig. \ref{fig:3dfit} is of about 360
simulations. Extending this work to an arbitrarily large number of EOSs
is computationally prohibitive.

However, we can exploit the existence of a quasi-universal behaviour of
$M_{\rm th}$
\citep{Bauswein2013,Koeppel2019,Agathos2019,Bauswein2020c}. More
specifically, we extend the quasi-universal relation derived by
\citet{Koeppel2019} by proposing that the functional dependence of
$\Mth({\rm EOS}, q, \chi)$ can be split into a part that is dependent on
the EOS and a part dependent on $q$ and $\chi$. From a mathematical point
of view, this essentially amounts to the separability in the functional
dependence and hence in adopting the following \textit{ansatz}
\begin{align}
  \Mth({\rm EOS},q,\chi) &= \kappa({\rm EOS})\,f(q, \chi) \,,
  && \label{eq:odeansatz} 
\end{align}
where the dependence on the EOS is expressed via a multiplicative
function following the study of \citet{Koeppel2019}
\begin{align}
\kappa({\rm EOS}) &:= \left(a-\frac{b}{1-c\,\mathcal{C}_{_{\rm
      TOV}}} \right) M_{_{\rm TOV}}\,. && \label{eq:Keos}
\end{align}
where $\mathcal{C}_{_{\rm TOV}}$ is the compactness of the nonrotating
stellar configuration with the maximum mass and ${R}_{_{\rm TOV}}$ its
radius, \ie $\mathcal{C}_{_{\rm TOV}}:={M}_{_{\rm TOV}}/{R}_{_{\rm
    TOV}}$. The coefficients in expression \eqref{eq:Keos} have been
reported by \citet{Koeppel2019} and are $a=2b/(2-c)$, $b=1.01$, $c =
1.34$.

In practice, expression \eqref{eq:odeansatz} proposes that $f(q, \chi)$
is a surface that models $\Mth$ as a function of $q$ and $\chi$
independently of the EOS. This surface can then be rescaled via a
function describing the EOS, $\kappa({\rm EOS})=\kappa(R_{\rm TOV},M_{\rm
  TOV})$, depending uniquely on the stellar compactness for the
maximum-mass nonrotating configuration. Stated differently, the
quasi-universal expression for $M_{\rm th}$ is expressed as $\hat{M}_{\rm
  th}:=\Mth/\kappa({\rm EOS})$, where all the dependence on the mass
ratio and spin of the binary is contained in the function $f(q,\chi)$,
whose behaviour remains to be determined.

\begin{figure}
  \centering
  \includegraphics[width=0.49\textwidth]{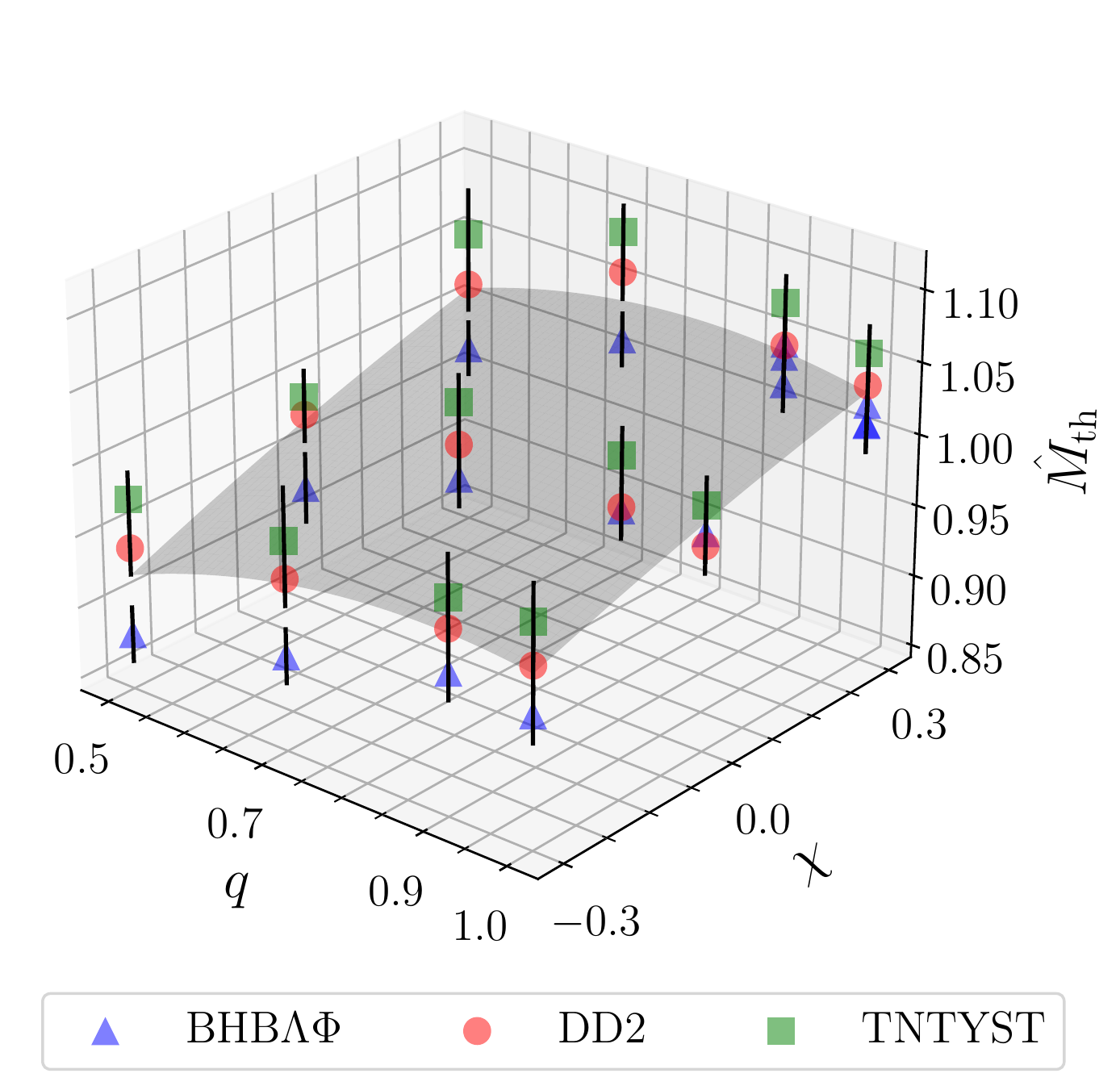}
  \caption{Similar to Fig. \ref{fig:3dfit} but highlighting the
    quasi-universal dependence of $\hat{M}_{\rm th}$ on the mass ratio
    and spin of the binary. Shown with different symbols are the values
    of $\hat{M}_{\rm th}$ and the corresponding uncertainties, while the
    gray-shaded surface represents the fit via \eqref{eq:odeansatz}.}
  \label{fig:univfit}
\end{figure}

Considering the nonlinear behaviour shown by the three fitting functions
in Fig. \ref{fig:3dfit}, it is natural to express the ansatz for
$f(q,\chi)$ via a second-order polynomial, \ie
\begin{align}
  f(q, \chi) :=\, &a_1 + a_2 (1-q) + a_3 \chi + a_4 (1-q) \chi + \nonumber\\
  &a_5 (1-q)^2 + a_6 \chi^2\,. &&
  \label{eq:f_ansatz}
\end{align}
Since we wish to recover the quasi-universal fit obtained by
\citet{Koeppel2019}, we set $a_1=1$ in \eqref{eq:f_ansatz}.

Note that the data clearly shows a non-monotonic growth of $M_{\rm th}$
as the mass asymmetry in the binary changes, which has been discussed
already by \citet{East2016} and \citet{Bauswein2020c} 
and more recently by \citet{Papenfort2021c}. 
Although the earlier works were restricted to
nonspinning binaries with larger mass ratios, the evidence for a
non-monotonic dependence remained tentative and, indeed, the fitting
expressions proposed by \citet{Bauswein2020c} are monotonic. However,
\citet{Papenfort2021c} has shown non-monotonic behavior should be expected 
due to an increase in accretion disk mass.

On the other hand, the presence of additional angular momentum in the
system stabilizes the remnant against gravitational collapse
\citep{Breu2016,Weih2017}. This implies that the fitting function $f$
should monotonically increase with spin. Since the maximum dimensionless
spin of uniformly rotating NSs is well constrained to be $\chi_{\rm
    max} \simeq 2\times0.65$, \citep[see, \eg the discussion
  in][]{Most2020d}, we can improve \eqref{eq:f_ansatz} by requiring that
$M_{\rm th}$ has a maximum for the maximum allowed value of the
dimensionless spin, \ie
\begin{equation}
  \label{eq:chimax}
  \left. \partial_{\chi} f(q,\chi) \right|_{\chi_{\rm max}}=0.
\end{equation}
Imposing \eqref{eq:chimax} on \eqref{eq:f_ansatz} leads to
\begin{align}
  a_6 &:= -\frac{a_3 + a_4 (1-q)}{2\, \chi_{\rm max}}\,, &&
\end{align}
thus leaving only four independent fitting coefficients, whose values are
$a_2=0.11, a_3=0.12, a_4=0.07, a_5=-0.3.$ We note that we
obtain essentially the same coefficients either when fitting all the
values of $\hat{M}_{\rm th}$ for the same values of $q$ and $\chi$ or
when averaging the coefficients obtained from the three distinct fits to
the different EOSs (see Tab. 3 of the AM for the coefficients of
  the three fits).

Figure \ref{fig:univfit} reports the fit of the quasi-universal threshold
mass for the combined data, showing the accurate representation of the
functional behaviour. Indeed, the fit yields a chi-squared of
$X^2=0.028$, with an average (maximum) deviation from the
  fit of only $\sim 2 \%$ ($\sim 6\%$). The very good match between the
data and the fitting function provides strong evidence for the existence
of the quasi-universal behaviour conjectured with the separability ansatz
\eqref{eq:odeansatz} and captures nicely the dependence of $M_{\rm th}$
on the mass ratio and spin of the binary. In particular, it highlights
that -- when compared to the irrotational case -- $M_{\rm th}$ increases
$\sim 6\%$ for the aligned-spin binaries considered here, and decreases
$\sim 10\%$ for the anti-aligned spin binaries. We note that while the
fit has considered all of the 40 binary configurations, we have verified
that predictions for $\Mth$ made when considering only $\chi_2 = 0$ match
equally well the numerical results obtained when $\chi_2$ is nonzero.

\begin{figure*}[t!]
  \centering
  \includegraphics[width=0.32\textwidth]{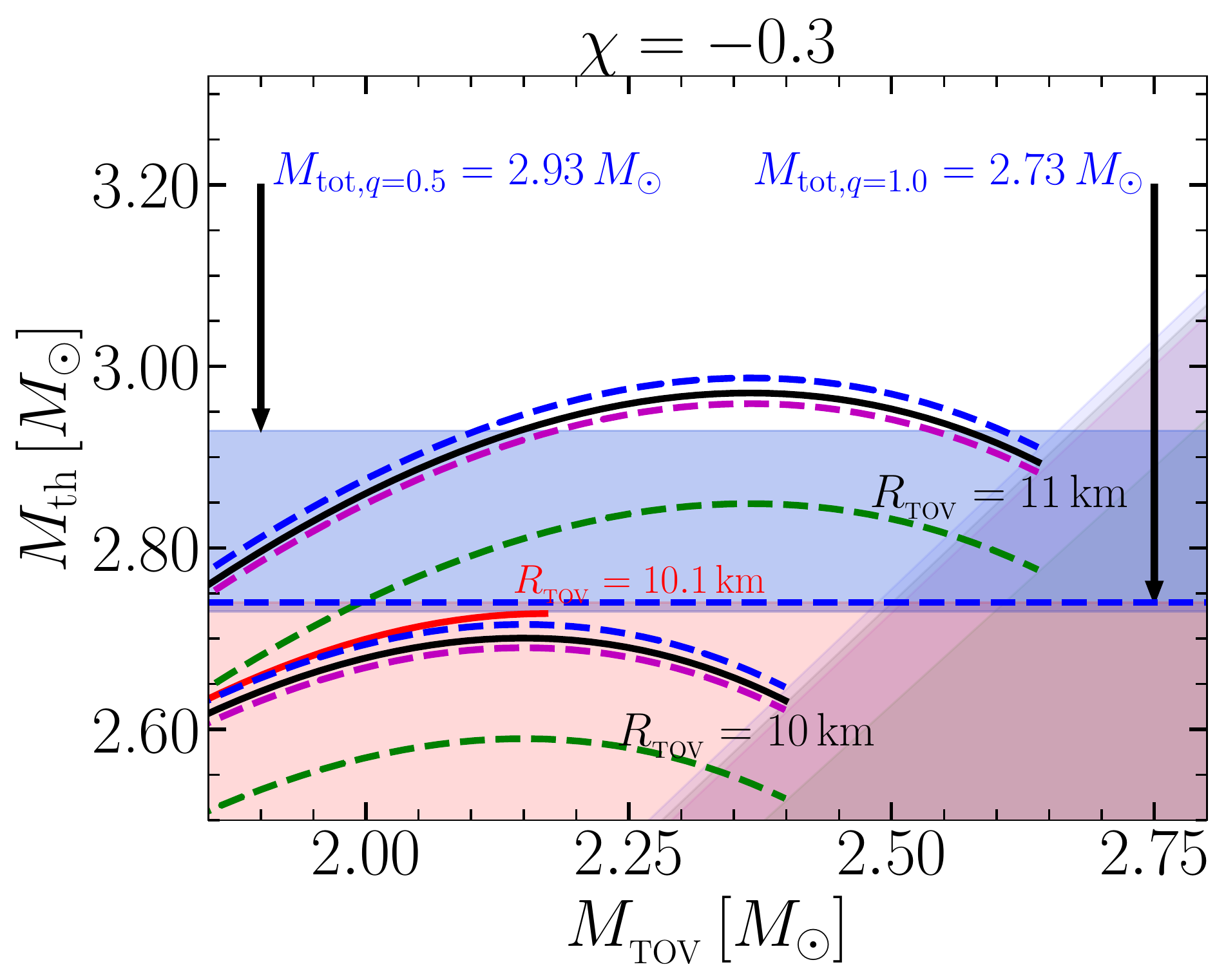}
  \includegraphics[width=0.32\textwidth]{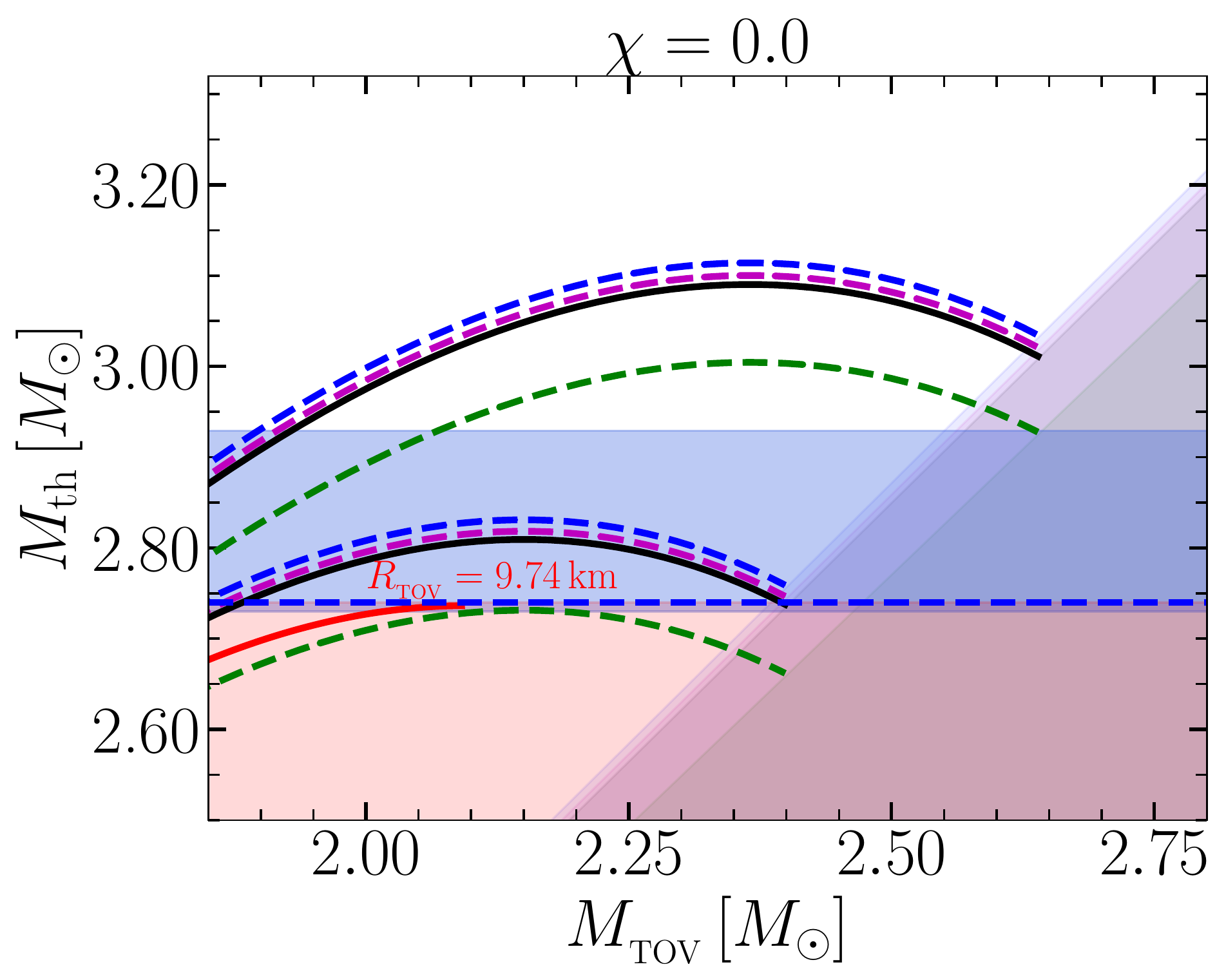}
  \includegraphics[width=0.32\textwidth]{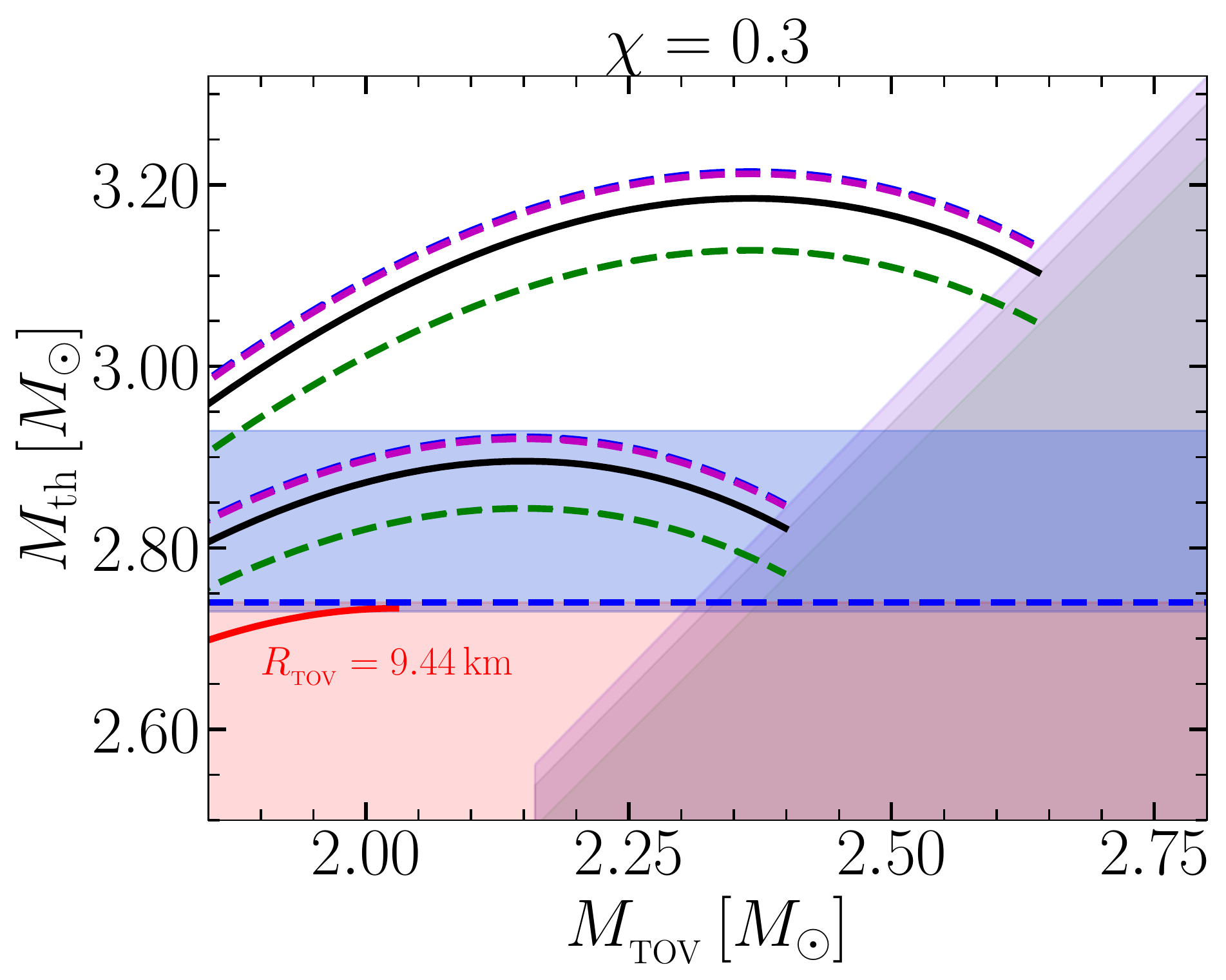} \\
  \vspace{0.25cm}
  \includegraphics[width=0.32\textwidth]{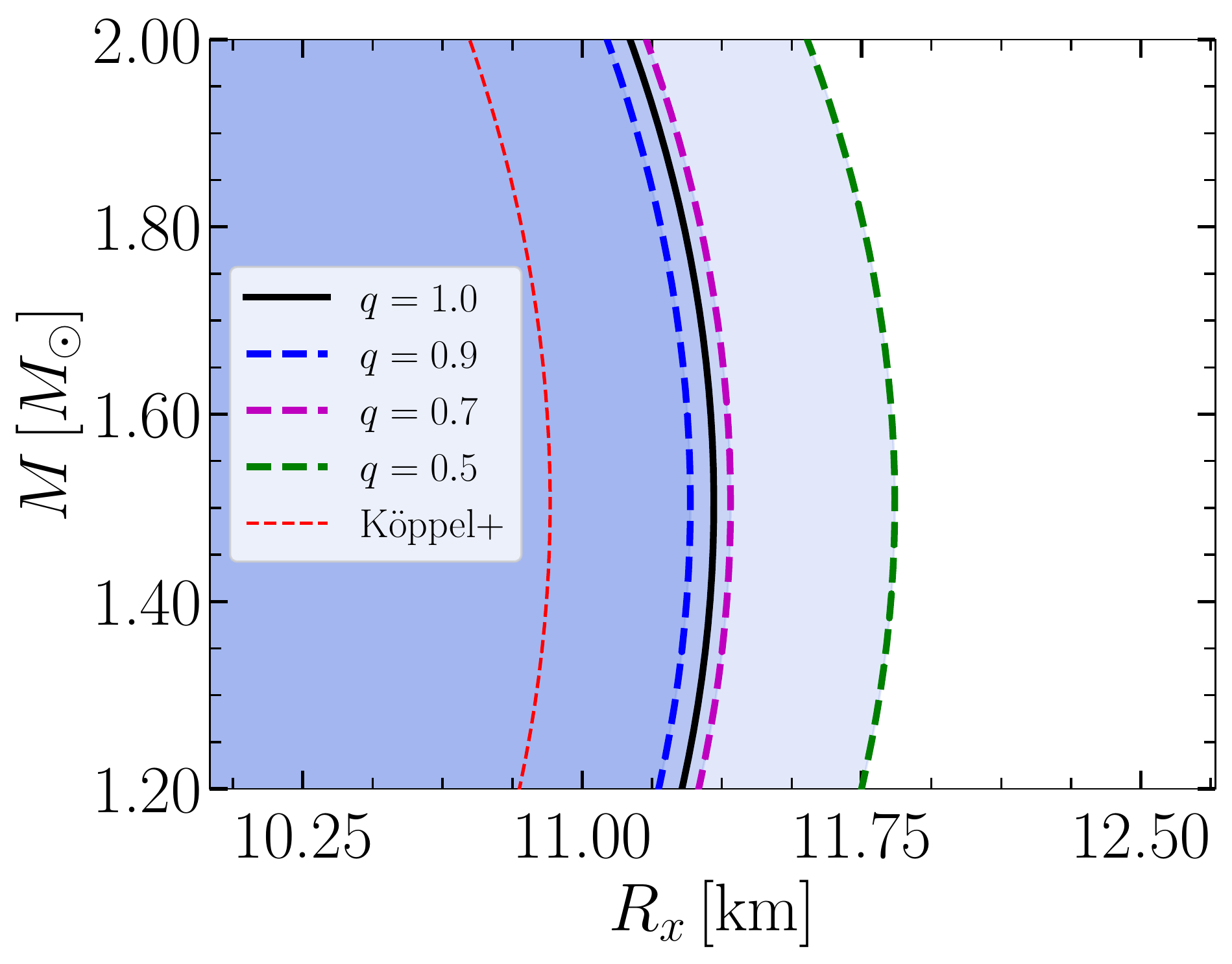}
  \includegraphics[width=0.32\textwidth]{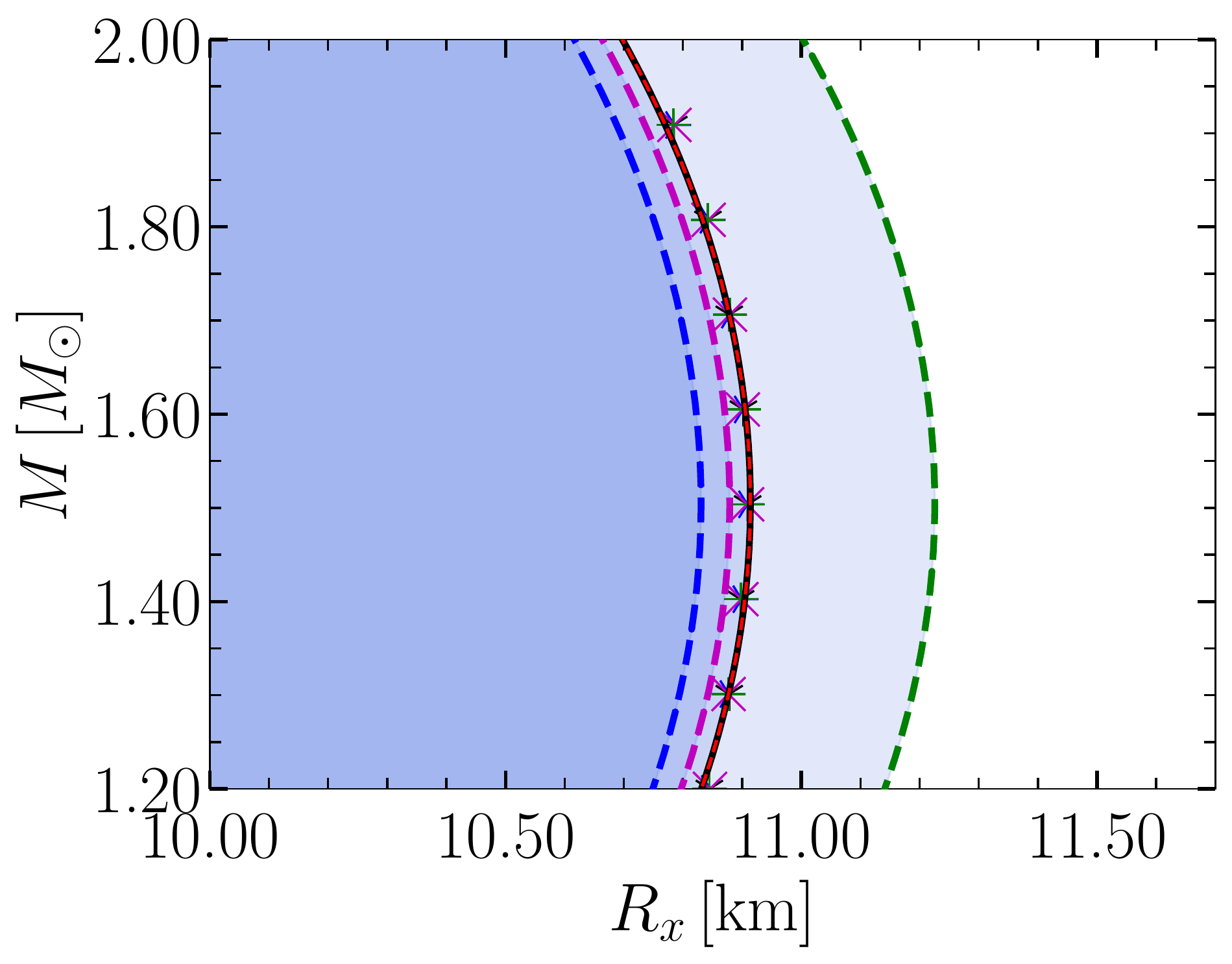}
  \includegraphics[width=0.32\textwidth]{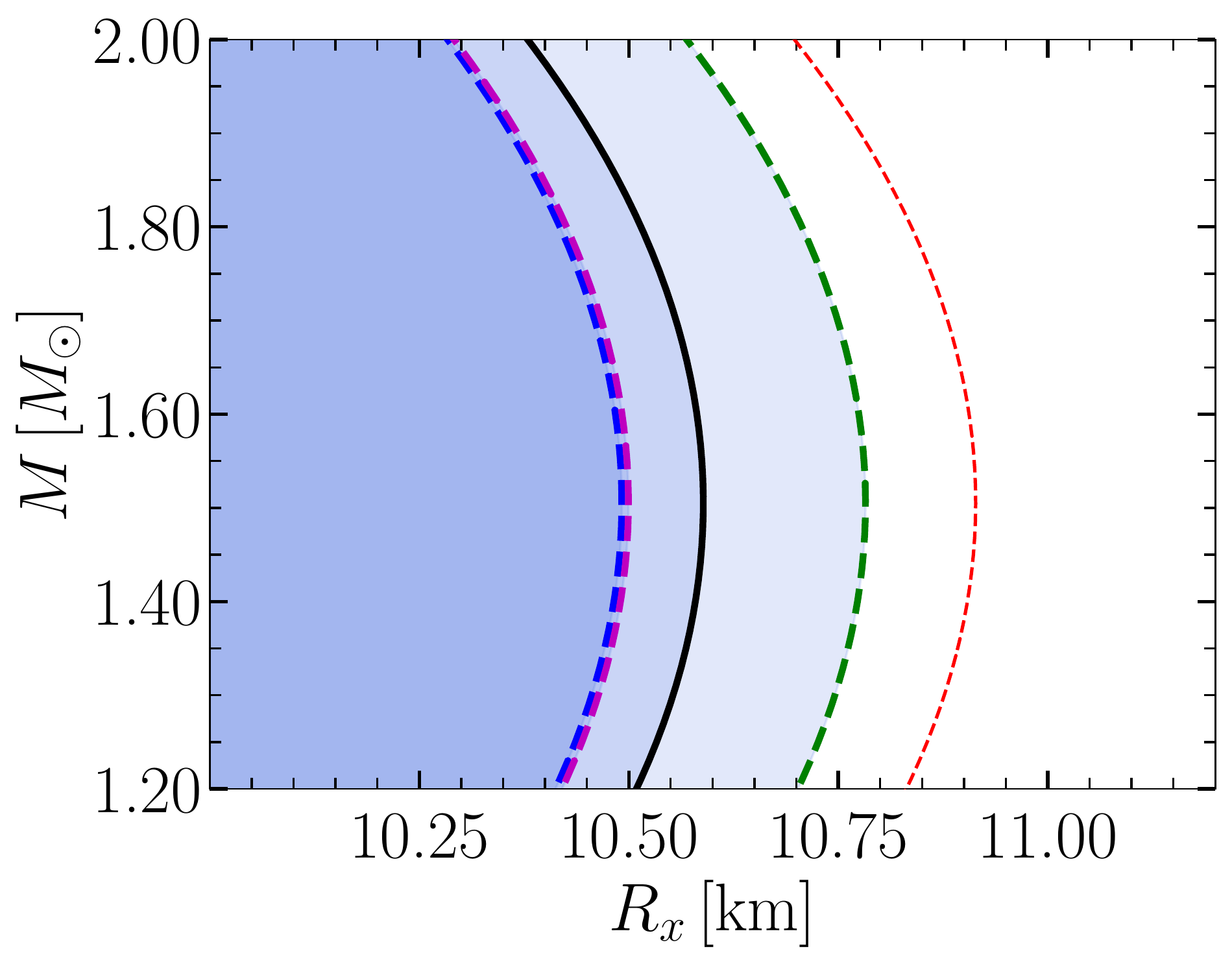}
  \caption{\textit{Top:} the lower bounds on $R_{_{\rm TOV},q=1}$ (red)
    using \textit{ansatz} \eqref{eq:odeansatz} for $R_{_{\rm TOV}}=10,
    11\, {\rm km}$. The black continuous line corresponds to $q=1$ while
    the coloured dashed lines mark the constraints set by
    $q=0.9\,,0.7\,,{\rm and}~0.5$. The blue region defines the mass of
    GW170817 and its uncertainty depending on $q$.  The red-shaded area
    shows the values excluded by the detection. The grey-shaded area
    represents values excluded by the causality constraint. Each panel,
    from left to right, corresponds to $\chi=[-0.3,0,0.3]$
    respectively. \textit{Bottom:} universal relation for the lower limit
    of $R_{x}(M,q,\chi_1)$. In the middle panel, we demonstrate $R_x(M)$
    is recovered when plotting $f(q,\chi)R_x(M,q,\chi)$. Overall, the top
    and bottom panels extend Fig. 4 of \citet{Koeppel2019} to binaries
    with unequal masses and nonzero spin.}
  \label{fig:radius_compare}
\end{figure*}

\noindent\textit{Lower Limits on the Stellar Radii.~} Following
\citet{Koeppel2019}, we can use Eqs. \eqref{eq:odeansatz} --
  \eqref{eq:f_ansatz}, to set lower limits on the radii of possible
  stellar models by computing sequences of $\Mth$ and $M_{_{\rm TOV}}$
  for fixed radii, as shown in Fig.~\ref{fig:radius_compare}. More
  specifically, we recall that once a value for $R_{_{\rm TOV},q=1}$ is
  fixed, expression \eqref{eq:odeansatz} selects a line in the
  $(\Mth,M_{_{\rm TOV}})$ plane. The first intersection of this line with
  the measured mass of a BNS with total mass $M_{\rm tot}$ that has not
  collapsed promptly sets a lower limit on $M_{\rm th}$
  \citep{Bauswein2017b}. This is shown in the upper panels of
Fig.~\ref{fig:radius_compare} [\cf left panel of Fig. 4 of
  \citet{Koeppel2019}] -- one for each of the spins considered here --
and where we report with a horizontal blue-dashed line the total
gravitational mass estimated for GW170817, $M_{{\rm
    tot},q=1}=2.74^{+0.04}_{-0.01}\,M_{\odot}$~\citep{Abbott2017}. Since
  the mass ratio is not well-known, the constraint for GW170817 is
  actually given by a band (blue-shaded area) whose vertical edges depend
  on the mass ratio $q$ (see arrows in the top-left panel of
  Fig. \ref{fig:radius_compare}). Also reported in the top panels of
Fig.~\ref{fig:radius_compare} with a grey-shaded area is the limit set by
causality and that requires $M_{_{\rm TOV}}/R_{_{\rm TOV}}\lesssim0.354$.

In essence, therefore, the blue band constrains the red-shaded
  area from above, yielding a lower limit for the radius of the
maximum-mass star, $R_{_{\rm TOV},q=1}$ (red solid line). The most
important difference with a similar figure presented by
\citet{Koeppel2019} for $q=1$ and $\chi=0$, is that we can now exploit
the dependence of $M_{\rm th}$ on the mass ratio (and spin) to report the
lower limit on $R_{_{\rm TOV}}$ for different values of $q$ (dashed
lines). In this way it is possible to appreciate that in the case of
anti-aligned spins (\eg for $\chi=-0.3$), very strong constraints can be
put on $R_{_{\rm TOV}, q=1}$ as the threshold mass is, in this case,
considerably smaller (\cf $R_{_{\rm TOV}, q=1}=10.24\,{\rm km}$ for
$\chi=-0.3$ vs $R_{_{\rm TOV}, q=1}=9.44\,{\rm km}$ for
$\chi=0.3$). Overall, the top panels in Fig. \ref{fig:radius_compare}
highlight how the knowledge of the mass ratio and spin of the binary can
be extremely powerful in setting lower limits on the stellar radii.

As in the analysis by \citet{Koeppel2019}, we can set lower
  limits not only for $R_{_{\rm TOV}}$, but for any arbitrary mass
  $M$. More specifically, we have found a fit for the minimum radius as a
  function of the mass ratio and spin in the binary, \ie $R_x(M, q,
  \chi)$, and verified that this function can be recovered by the
  original expression $R_x(M)$ in \citet{Koeppel2019} for $q=1$ and
  $\chi=0$ (see symbols in the lower-middle panel of
  Fig.~\ref{fig:radius_compare})
\begin{align}
  R_x(M)=-0.88\, M^2 + 2.66\, M + 8.91\,,
\end{align}
via the same scaling function $f(q,\chi)$ in
  Eq. \eqref{eq:f_ansatz}, namely
\begin{align}
  R_x(M, q, \chi)=\frac{R_x(M)}{f(q, \chi)}\,.
  \label{eq:Rx:ansatz}
\end{align}
Because relation \eqref{eq:Rx:ansatz} should be seen as
  an ansatz representative of the EOSs used here, it might need
additional corrections for EOSs that include strong phase transitions
\citep{Most2018b,Weih2020,Bauswein2020}, or that include regimes
  of extremely high densities \citep{Most2021c}.

The bottom panels of Fig. \ref{fig:radius_compare} [\cf right panel of
  Fig. 4 of \citet{Koeppel2019}] report the estimates of $R_x(M, q,
\chi)$ for different values of $q$ (different dashed lines) and $\chi$
(different columns); in each plot, we include the original fit for $R_x$
from \citet{Koeppel2019} as a reference (red dashed line). Note that, for
a given spin of the binary, the knowledge of the mass ratio can
considerably increase the lower limit on the stellar radii, especially
for systems with negative total spin. At the same time, given the
non-monotonic nature of the dependence on $q$, equal-mass systems do not
necessarily yield the weakest constraint, which is instead attained for
$q\simeq 0.8-0.9$.

\section{Conclusions}
\label{sec:conclusions}

We have performed the first systematic study of the impact that mass
asymmetry and spin have on the threshold mass of BNS systems, $M_{\rm
th}$. We have done so by measuring $\Mth$ for 40 different BNS configurations
encompassing three temperature dependent EOSs, four mass ratios, and
a systematic sampling of spin configurations.
Using this data we have derived a quasi-universal relation for $M_{\rm
th}$ and expressed its dependence on the mass ratio and spin of the
binary. The new expression recovers the results of \citet{Koeppel2019}
for equal-mass, irrotational binaries, and reveals that $M_{\rm th}$ can
increase (decrease) by $5\%~(10\%)$ for binaries that have spins aligned
(anti-aligned) with the orbital angular momentum. In addition, we find
evidence for a non-monotonic dependence of $M_{\rm th}$ on the mass
asymmetry in the system, which can be explained by an increase in accretion disk mass
\citep{Papenfort2021c}. Furthermore, we have extended to unequal-masses
and spinning binaries the lower limits that can be set on the stellar
radii once a neutron-star binary is detected, obtaining generic and
analytic expressions for the minimum radii as a function of the mass
ratio and of the total spin of the binary. In this way, we have
highlighted that the merger of an unequal-mass, rapidly spinning binary
can significantly constrain the allowed values of the stellar radii.

With more than 360 binaries simulated, this work has represented a
challenge both for the associated computational costs and for the amount
of data to be analysed. Furthermore, while it has provided a first sparse
but complete survey of the space of parameters to be expected in BNSs, it
can be improved in a number of ways. First, by investigating more
carefully the role played by spin of the secondary star. Second, by
increasing the number of EOSs considered so -- especially if they involve
a softening from phase transitions \citep{Most2018b,Weih2020} -- as to
further refine the properties of the quasi-universal behaviour. Third, by
elucidating the role played by the disruption of the stars at
merger. Finally, by exploring the possibility that ultra-strong magnetic
fields could modify the stability properties of the merged object and
hence the threshold mass. We leave all of these improvements to future
works.

\section*{Acknowledgements}

It is a pleasure to thank C. Ecker for useful discussions. LR
  acknowledges funding from the Hessian Research Cluster ``ELEMENTS'' and
  from the ERC Advanced Grant ``JETSET'' (Grant No. 884631). ERM
acknowledges support from the Princeton Center for Theoretical Science,
the Princeton Gravity Initiative, and the Institute for Advanced
Study. The simulations were performed on HPE Apollo Hawk at the High
Performance Computing Center Stuttgart (HLRS) under the grants BBHDISKS
and BNSMIC, and on SuperMUC at the Leibniz Supercomputing Centre.

\software{Einstein Toolkit \citep{loeffler_2011_et},
          \texttt{Carpet} \citep{Schnetter-etal-03b},
          \texttt{FIL} \citep{Most2019b, Etienne2015}, \texttt{FUKA}
          \citep{Papenfort2021b}, \texttt{Kadath} \citep{Grandclement09}
}

\bibliography{paper.bbl}

\newpage
\appendix
\begin{deluxetable*}{|l|c|c|c|c|c|c|c|c|c|c|c|}
  \centerwidetable \tablewidth{0pt} \tablecaption{Lists of the binary
    configurations explored in this study for the three nuclear EOSs
    considered. Included is the determined threshold mass, $M_{\rm th}$;
    the corresponding chirp mass, $\mathcal{M}_{\rm th}$; the ADM mass,
    $M_{_{\rm ADM}}$, of each NS at infinite separation; the
    dimensionless spin of the each NS, $\chi_1 \,, \chi_2$; and the effective spin of
    the binary, $\tilde{\chi}$. In all cases, the spin axis is parallel 
    to the orbital rotation axis.
    $\Mth$ in configurations with a $\texttt{*}$ were obtained using the
    \textit{averaging method} as discussed in the Appendix
    Materials\label{tab:configs}}
  \tablehead{ & $M_{\rm th}$ & $\mathcal{M}_{\rm th}$ & $M_{1}$ & $M_{2}$
    & $q$ & $\chi_{_{\rm 1}}$ & $\chi_{_{\rm 2}}$ & $\tilde{\chi}$ & EOS \\
    & $\left[M_{_{\rm TOV}}\right]$ & $\left[M_\odot\right]$  & $\left[M_\odot\right]$ &
      $\left[M_\odot\right]$     & $\left[M_\odot\right]$  & & & & 
   }
  \startdata
  \texttt{BHB}$\Lambda\Phi$\texttt{.1.0.-0.3}  & 2.961 & 1.289 & 1.481 & 1.481 & 1.0 & -0.3 & ~0.0 & -0.150 & \texttt{BHB}$\Lambda\Phi$ \\
  \texttt{BHB}$\Lambda\Phi$\texttt{.1.0.+0.0}  & 3.149 & 1.370 & 1.574 & 1.574 & 1.0 & ~0.0 & ~0.0 & ~0.000 & \texttt{BHB}$\Lambda\Phi$ \\
  \texttt{BHB}$\Lambda\Phi$\texttt{.1.0.+0.2}  & 3.197 & 1.392 & 1.599 & 1.599 & 1.0 & ~0.2 & ~0.1 & ~0.150 & \texttt{BHB}$\Lambda\Phi$ \\
  \texttt{BHB}$\Lambda\Phi$\texttt{.1.0.+0.3}  & 3.199 & 1.392 & 1.600 & 1.600 & 1.0 & ~0.3 & ~0.0 & ~0.150 & \texttt{BHB}$\Lambda\Phi$ \\
  \texttt{BHB}$\Lambda\Phi$\texttt{.1.0.+0.4}  & 3.242 & 1.411 & 1.621 & 1.621 & 1.0 & ~0.4 & -0.1 & ~0.150 & \texttt{BHB}$\Lambda\Phi$ \\
  \texttt{BHB}$\Lambda\Phi$\texttt{.0.9.-0.3}  & 2.982 & 1.296 & 1.569 & 1.412 & 0.9 & -0.3 & ~0.0 & -0.158 & \texttt{BHB}$\Lambda\Phi$ \\
  \texttt{BHB}$\Lambda\Phi$\texttt{.0.9.+0.0}  & 3.134 & 1.362 & 1.649 & 1.484 & 0.9 & ~0.0 & ~0.0 & ~0.000 & \texttt{BHB}$\Lambda\Phi$ \\
  \texttt{BHB}$\Lambda\Phi$\texttt{.0.9.+0.2}  & 3.227 & 1.402 & 1.699 & 1.529 & 0.9 & ~0.2 & ~0.1 & ~0.158 & \texttt{BHB}$\Lambda\Phi$ \\
  \texttt{BHB}$\Lambda\Phi$\texttt{.0.9.+0.3}  & 3.288 & 1.429 & 1.731 & 1.558 & 0.9 & ~0.3 & ~0.0 & ~0.158 & \texttt{BHB}$\Lambda\Phi$ \\
  \texttt{BHB}$\Lambda\Phi$\texttt{.0.9.+0.4}  & 3.317 & 1.442 & 1.746 & 1.571 & 0.9 & ~0.4 & -0.1 & ~0.158 & \texttt{BHB}$\Lambda\Phi$ \\
  \texttt{BHB}$\Lambda\Phi$\texttt{.0.7.-0.3}  & 2.879 & 1.230 & 1.694 & 1.185 & 0.7 & -0.3 & ~0.0 & -0.176 & \texttt{BHB}$\Lambda\Phi$ \\
  \texttt{BHB}$\Lambda\Phi$\texttt{.0.7.+0.0}  & 3.077 & 1.314 & 1.810 & 1.267 & 0.7 & ~0.0 & ~0.0 & ~0.000 & \texttt{BHB}$\Lambda\Phi$ \\
  \texttt{BHB}$\Lambda\Phi$\texttt{.0.7.+0.3}  & 3.209 & 1.371 & 1.888 & 1.321 & 0.7 & ~0.3 & ~0.0 & ~0.176 & \texttt{BHB}$\Lambda\Phi$ \\
  \texttt{BHB}$\Lambda\Phi$\texttt{.0.5.-0.3}  & 2.791 & 1.132 & 1.861 & 0.930 & 0.5 & -0.3 & ~0.0 & -0.200 & \texttt{BHB}$\Lambda\Phi$ \\
  \texttt{BHB}$\Lambda\Phi$\texttt{.0.5.+0.0}  & 2.929 & 1.188 & 1.952 & 0.976 & 0.5 & ~0.0 & ~0.0 & ~0.000 & \texttt{BHB}$\Lambda\Phi$ \\
  \texttt{BHB}$\Lambda\Phi$\texttt{.0.5.+0.3}  & 3.071 & 1.246 & 2.048 & 1.024 & 0.5 & ~0.3 & ~0.0 & ~0.200 & \texttt{BHB}$\Lambda\Phi$ \\
  \hline
  \texttt{DD2.1.0.-0.3}  & 3.209 & 1.397 & 1.605 & 1.605 & 1.0 & -0.3 & ~0.0 & -0.150 & \texttt{DD2} \\
  \texttt{DD2.1.0.+0.0}  & 3.266 & 1.421 & 1.633 & 1.633 & 1.0 & ~0.0 & ~0.0 & ~0.000 & \texttt{DD2} \\
  \texttt{DD2.1.0.+0.3}  & 3.437 & 1.496 & 1.718 & 1.718 & 1.0 & ~0.3 & ~0.0 & ~0.150 & \texttt{DD2} \\
  \texttt{DD2.0.9.-0.3}  & 3.219 & 1.399 & 1.694 & 1.525 & 0.9 & -0.3 & ~0.0 & -0.158 & \texttt{DD2} \\
  \texttt{DD2.0.9.+0.0}  & 3.287 & 1.428 & 1.730 & 1.557 & 0.9 & ~0.0 & ~0.0 & ~0.000 & \texttt{DD2} \\
  \texttt{DD2.0.9.+0.3}  & 3.467 & 1.507 & 1.825 & 1.642 & 0.9 & ~0.3 & ~0.0 & ~0.158 & \texttt{DD2} \\
  \texttt{DD2.0.7.-0.3}  & 3.190 & 1.362 & 1.877 & 1.314 & 0.7 & -0.3 & ~0.0 & -0.176 & \texttt{DD2} \\
  \texttt{DD2.0.7.+0.0}  & 3.299 & 1.409 & 1.940 & 1.358 & 0.7 & ~0.0 & ~0.0 & ~0.000 & \texttt{DD2} \\
  \texttt{DD2.0.7.+0.3}  & 3.517 & 1.502 & 2.069 & 1.448 & 0.7 & ~0.3 & ~0.0 & ~0.176 & \texttt{DD2} \\
  \texttt{DD2.0.5.-0.3}  & 3.123 & 1.267 & 2.082 & 1.041 & 0.5 & -0.3 & ~0.0 & -0.200 & \texttt{DD2} \\
  \texttt{DD2.0.5.+0.0}  & 3.240 & 1.314 & 2.160 & 1.080 & 0.5 & ~0.0 & ~0.0 & ~0.000 & \texttt{DD2} \\
  \texttt{DD2.0.5.+0.3}  & 3.369 & 1.366 & 2.246 & 1.123 & 0.5 & ~0.3 & ~0.0 & ~0.200 & \texttt{DD2} \\
  \hline
  \texttt{TNTYST.1.0.-0.3*} & 2.870 & 1.249 & 1.435 & 1.435 & 1.0 & -0.3 & ~0.0 & -0.150 & \texttt{TNTYST} \\
  \texttt{TNTYST.1.0.+0.0}  & 2.916 & 1.269 & 1.458 & 1.458 & 1.0 & ~0.0 & ~0.0 & ~0.000 & \texttt{TNTYST} \\
  \texttt{TNTYST.1.0.+0.3}  & 3.048 & 1.327 & 1.524 & 1.524 & 1.0 & ~0.3 & ~0.0 & ~0.150 & \texttt{TNTYST} \\
  \texttt{TNTYST.0.9.-0.3*} & 2.855 & 1.241 & 1.503 & 1.352 & 0.9 & -0.3 & ~0.0 & -0.158 & \texttt{TNTYST} \\
  \texttt{TNTYST.0.9.+0.0}  & 2.956 & 1.284 & 1.556 & 1.400 & 0.9 & ~0.0 & ~0.0 & ~0.000 & \texttt{TNTYST} \\
  \texttt{TNTYST.0.9.+0.3}  & 3.095 & 1.345 & 1.629 & 1.466 & 0.9 & ~0.3 & ~0.0 & ~0.158 & \texttt{TNTYST} \\
  \texttt{TNTYST.0.7.-0.3*} & 2.845 & 1.215 & 1.674 & 1.171 & 0.7 & -0.3 & ~0.0 & -0.176 & \texttt{TNTYST} \\
  \texttt{TNTYST.0.7.+0.0}  & 2.950 & 1.260 & 1.735 & 1.214 & 0.7 & ~0.0 & ~0.0 & ~0.000 & \texttt{TNTYST} \\
  \texttt{TNTYST.0.7.+0.3}  & 3.135 & 1.339 & 1.844 & 1.290 & 0.7 & ~0.3 & ~0.0 & ~0.176 & \texttt{TNTYST} \\
  \texttt{TNTYST.0.5.-0.3}  & 2.810 & 1.139 & 1.873 & 0.936 & 0.5 & -0.3 & ~0.0 & -0.200 & \texttt{TNTYST} \\
  \texttt{TNTYST.0.5.+0.0}  & 2.850 & 1.156 & 1.900 & 0.950 & 0.5 & ~0.0 & ~0.0 & ~0.000 & \texttt{TNTYST} \\
  \texttt{TNTYST.0.5.+0.3*} & 3.030 & 1.229 & 2.020 & 1.010 & 0.5 & ~0.3 & ~0.0 & ~0.200 & \texttt{TNTYST} \\
  \enddata
\end{deluxetable*}

\section{Numerical Methods and Setup}
\label{sec:methods}

\noindent\textit{Numerical Simulations.~} As mentioned in the main text,
our ID is constructed making use of the recently developed initial-data
solver \texttt{FUKA}, which is based on the \texttt{KADATH} spectral
solver library, solving the eXtended Conformal Thin Sandwich (XCTS)
formulation of Einstein's field equations
\citep{Pfeiffer:2002iy,Pfeiffer:2005,Papenfort2021b}. The ID is initially
constructed using the force-balance equations to determine a
quasi-circular orbit. The spin of each NS is achieved by modeling the
velocity field of the fluid as a linear combination of a purely
irrotational component and a uniformly rotating component
\citep{Tacik15,Tsokaros2015,Papenfort2021b}. To minimize the eccentricity
of the inspiral, we utilize $3.5$ Post-Newtonian order estimates of the
expansion factor, $\dot{a}$, and the orbital velocity as discussed in
\citet{Papenfort2021b}. This is important as quasi-circular ID for
asymmetric binaries and binaries with spinning objects result in very
eccentric inspirals which is also discussed in \citet{Papenfort2021b}. Additionally,
it has been shown that eccentricity can have an impact on the stability
of the remnant\citep{East2016}, therefore, the use of 3.5 PN estimates
is important to obtain accurate measurements of $\Mth$. 
Finally, the initial separation of the compact objects are set to
$50 \rm km$, thus allowing the binary to equilibrate over the course of a
few orbits prior to merger due to the approximations used in the initial
data construction.

\noindent\textit{Realistic, hot EOSs.~}
The binaries simulated here have been modelled with three different
realistic and temperature-dependent (hot) EOSs. In particular, they are
the rather soft \texttt{TNTYST} EOS \citet{Togashi2017}, the rather stiff
\texttt{BHB}$\Lambda\Phi$ EOS \citet{Banik2014}, and the intermediate
(HS-)DD2 EOS \citep{Typel2010}. The most salient properties of these
EOSs, namely, the maximum mass of the nonrotating configuration, the
corresponding radius, the compactness, and the free-fall timescales are
reported in Tab. \ref{tab:eoss}.

\begin{deluxetable}{l|cccc}[b!]
  \tablecaption{For the EOss considered here we report the maximum mass
    of a nonrotating configuration, $M_{_{\rm TOV}}$, together with the
    corresponding radius, $R_{_{\rm TOV}}$, compactness, $C_{_{\rm
        TOV}}$, and the free-fall timescale $\tau_{_{\rm
        TOV}}$. \label{tab:eoss}}
  \tablehead{
    & $M_{_{_{\rm TOV}}}$ & $R_{_{_{\rm TOV}}}$ & $C{_{_{\rm TOV}}}$ & $\tau_{_{\rm TOV}}$ \\
    & $[M_{\odot}]$  & $[{\rm km}]$  &                & $[\mu {\rm s}]$ }
  \startdata
  \texttt{BHB}$\Lambda\Phi$ & 2.10 & 11.64 & 0.26 & 83.31 \\
  \texttt{DD2}              & 2.42 & 11.94 & 0.30 & 80.60 \\
  \texttt{TNTYST}           & 2.23 & 10.17 & 0.32 & 66.12
  \enddata
\end{deluxetable}

\noindent\textit{Spin Configurations.~} Emphasis in our analysis has been
placed on low-spin priors and on the role of the spin of the primary in
determining the threshold mass. Hence, the large majority of our
binaries has an irrotational secondary (\ie $\chi_2=0$). However,
binaries with non-negligible mixed spins cannot be ruled out and, thus,
we have performed additional spot tests which include configurations of
constant mass ratio and effective spin, but with a spinning secondary so
as to ascertain the difference in the collapse time using the
\texttt{BHB}$\Lambda\Phi$ EOS. All of the 40 binaries
considered here and their properties are collected in
Tab. \ref{tab:configs}.

With the data collected in this way we have performed a double
analysis. First, we have considered only binaries with irrotational
secondaries and performed the fits as discussed in the main text.  Next,
using the resulting coefficients we have compared the predictions of the
fit with the actual numerical results of the four binaries with spinning
secondary. In this way, we have found that the analytic predictions and
the numerical results differ by less than $3 \%$ and
are therefore well within the average error of the fit. Overall, this
result has given us confidence that the spin of the secondary has a
negligible impact on $\Mth$, which is instead dynamically dominated by
the spin of the primary. This is a reasonable estimate for two distinct
reasons. Firstly, $\Mth$ effectively measures the strength of the
gravitational field of the merged object and is therefore dominated the
properties of the primary. Secondly, the modifications on $\Mth$ due to
spin are, overall, not large.  As a second approach, we have performed a
fit using all of the data available and this is what is actually
presented in the main text; the corresponding values of the fitting
coefficients and their uncertainties are collected in
Tab. \ref{tab:fits}.

\section{Comparison of Threshold Mass Measurements}
To determine $M_{\rm th}$ from our numerical simulations we have elected
to utilize the method of measuring the collapse time, $t_{\rm col}$, as
proposed in \cite{Koeppel2019} given the physical motivation behind it.
We will reference this as the \textit{``free-fall''}, method which is
discussed below. However, similar measurements can be made with minimal
impact to the result of the measured value of $\Mth$ by using a simpler
approach, which we refer to as the \textit{``averaging''} method, and
which averages the masses of the two binaries closest to the critical
value leading to a prompt collapse.

\noindent\textit{Free-Fall Method.~} The original method of measuring the
collapse time, $t_{\rm col}$, as proposed in \cite{Koeppel2019}, is
defined by tracking the minimum of the lapse, $\tilde{\alpha} :=
\min(\alpha)$. While this approach is adequate for equal-mass,
nonspinning binaries, it leads to potential biases when considering
unequal-mass and spinning binaries, since the lapse is not only dependent
on the lengthscales of the system, but it is strongly impacted by
asymmetries in the system. Therefore, when performing measurements of the
collapse time, $t_{\rm col}$, we analyze $\tilde{\alpha}$ when normalized
by its maximum, which we define as $\hat{\alpha} :=
{\tilde{\alpha}}/{\max(\tilde{\alpha})}$.

Because of this rescaling, new brackets must be applied to determine the
time of merger, $t_{\rm merge}$, and the time of black-hole formation,
$t_{_{\rm BH}}$. To do so, we set as the merger time the coordinate time
such that
\begin{align}
  &&t_{\rm merg}:& \qquad \hat{\alpha}_{\rm merge}=0.9\,.&
\end{align}
Similarly, we set as the threshold for black-hole formation and the
related coordinate time to be
\begin{align}
  &&t_{_{\rm BH}}:& \qquad \hat{\alpha}_{\rm BH}=0.1\,,&
\end{align}
such that the remnant is gravitationally unstable across all EOSs and
configurations considered and the formation of a BH is certain. Note that
the merged object never collapse faster than the shortest free-fall
timescale ($t_{\rm col} >= \tau_{\rm ff}$) where the free-fall timescale
is defined as \citep{Rezzolla_book:2013}
\begin{align}
  \tau_{\rm ff}(M,R) := \frac{\pi}{2} \sqrt{\frac{R^3}{2M}}\,,
\end{align}
and the shortest $\tau_{\rm ff}$ occurs at $\tau_{_{\rm TOV}} :=
\tau_{\rm ff}(M_{_{\rm TOV}}, R_{_{\rm TOV}})$ and is a physically
well-motivated lower limit. We recognise that both of these markers are
somewhat arbitrary, but, as done in \citet{Koeppel2019}, they reproduce
well the behaviour of the binaries when other markers, \eg the time of
the maximum of the gravitational-wave emission and the time of the
appearance of a black-hole ringdown, are used.

\begin{figure}[b!]
  \centering
  \includegraphics[width=0.49\columnwidth]{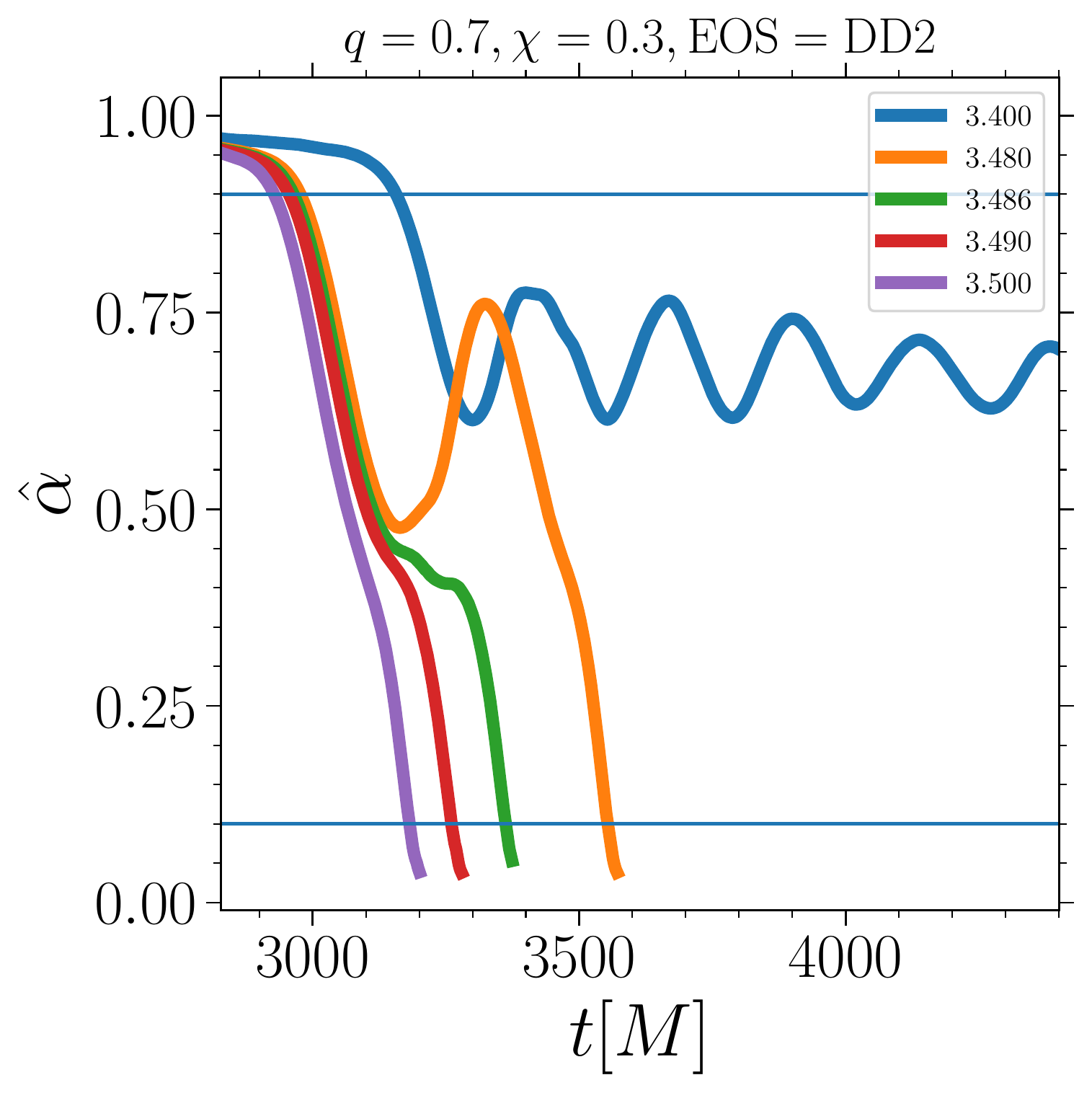}
  \includegraphics[width=0.49\columnwidth]{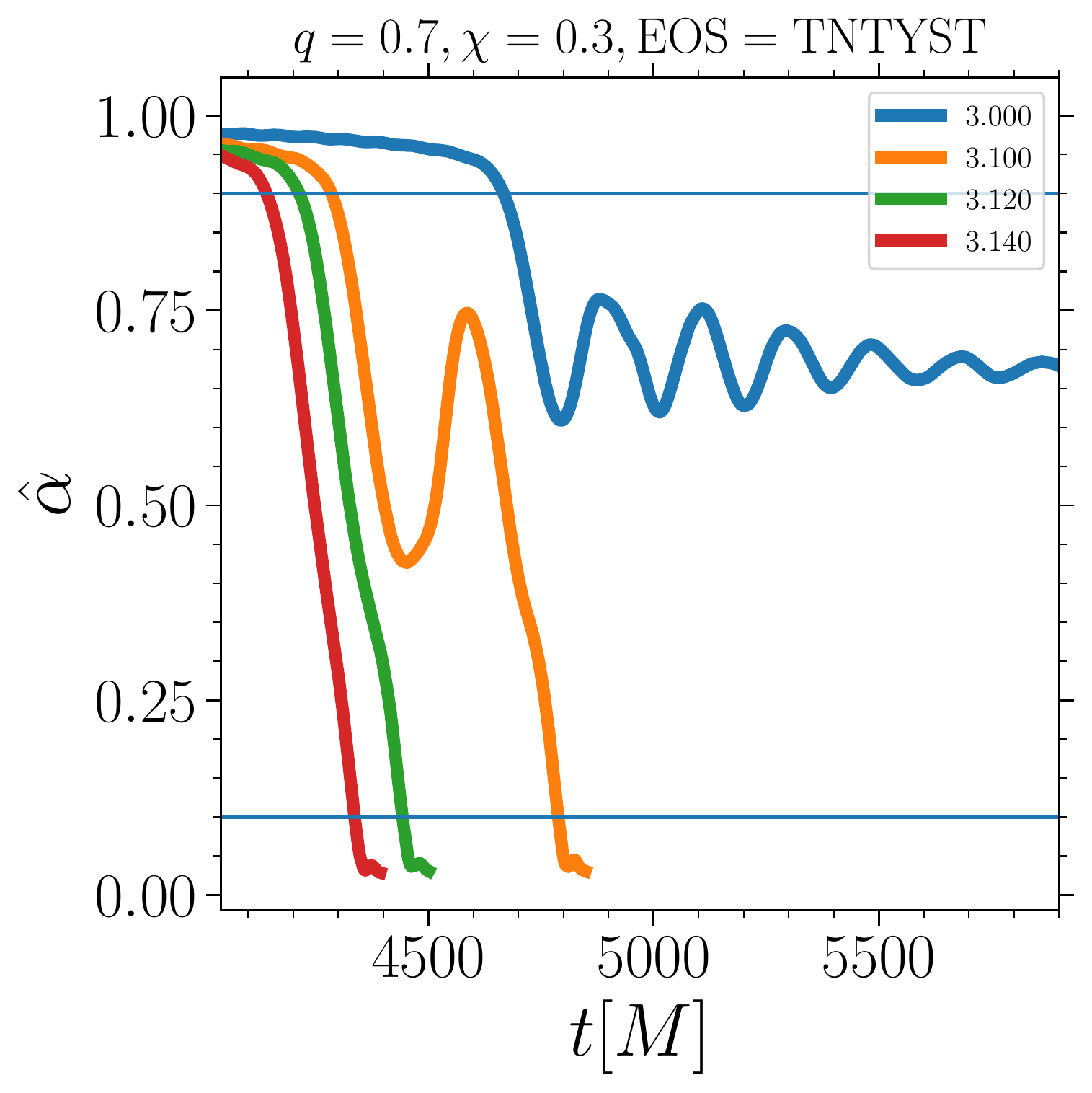} \\
  \includegraphics[width=0.49\columnwidth]{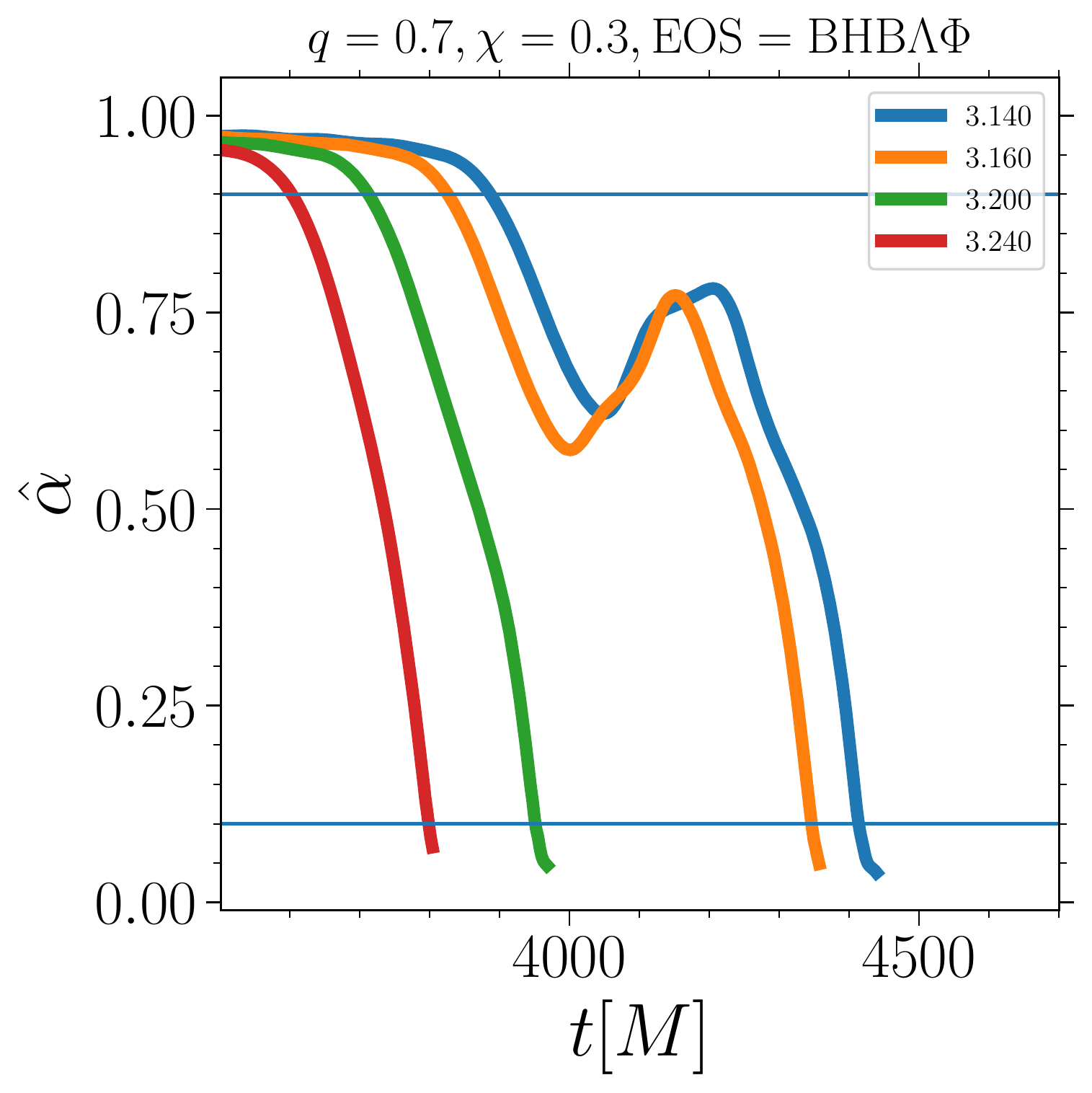} 
  \includegraphics[width=0.49\columnwidth]{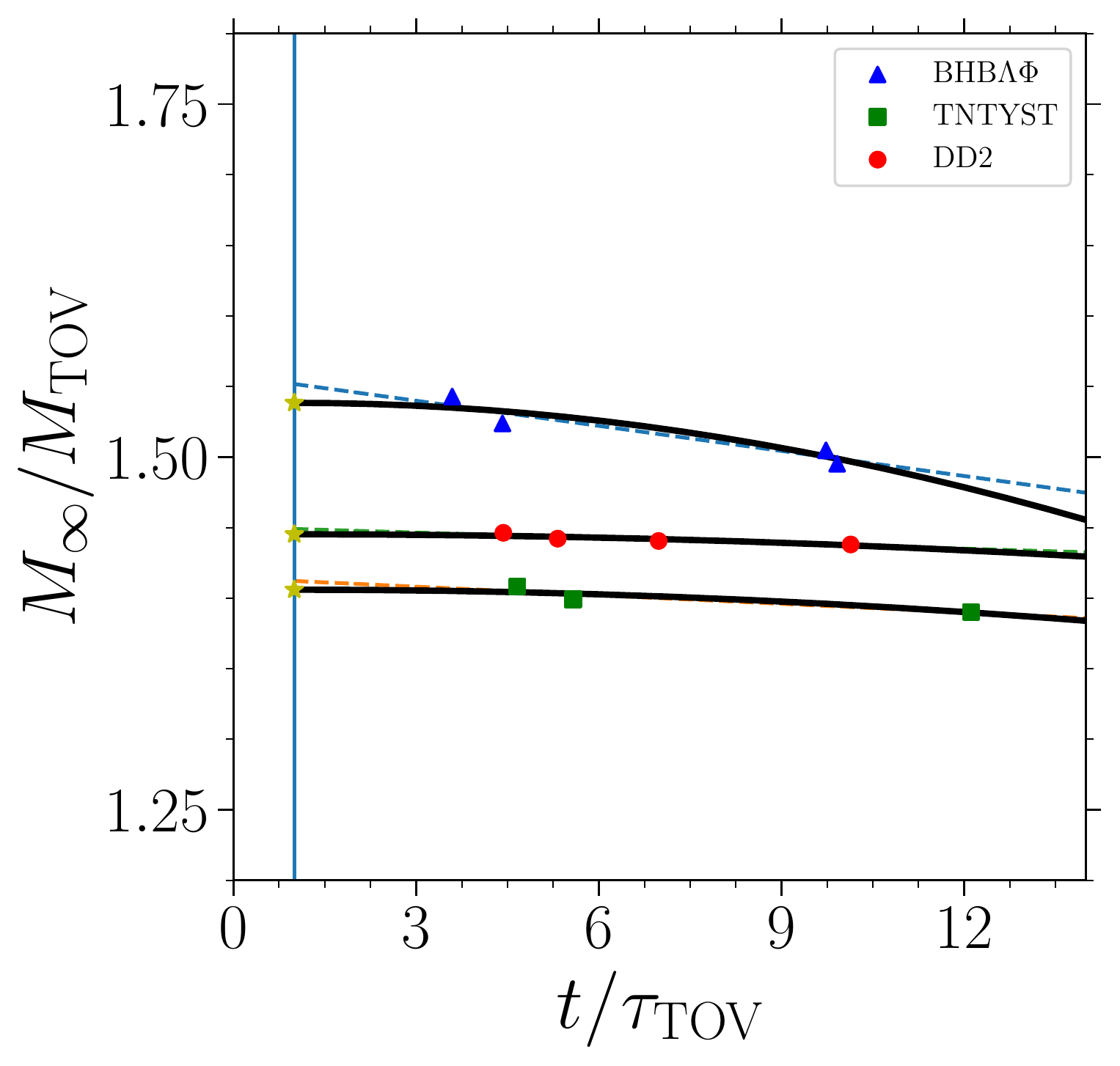}
  \caption{\textit{Top \& Bottom Left:} Plots of normalized minimum
    lapse, $\hat{\alpha}$, for $q=0.7$ and $\chi=0.3$ for each of the
    EOSs used in this work. Horizontal lines mark the threshold values to
    define merger and black-hole formation, \ie $\hat{\alpha}_{\rm
      merge}=0.9$ and $\hat{\alpha}_{\rm BH}=0.1$, respectively.
    \textit{Bottom Right:} Example of measuring $\Mth$ using the
    \textit{`` free-fall''} method for the three EOSs; dashed lines
    denote a linear-fit result.}
  \label{fig:lapse:compare}
\end{figure}

Using these markers, we measure the collapse time as defined as
\begin{align}
  t_{\rm col} := t_{_{\rm BH}} - t_{\rm merg}\,,
\end{align}
which is the same as the prescription from \cite{Koeppel2019}, but here
it is calculated based on $\hat{\alpha}$ and with different bracket
values.  In Fig. \ref{fig:lapse:compare} we show examples of how
$\hat{\alpha}$ vs coordinate time looks for all equations with $q=0.7 \,,
\chi_1 = 0.3 \,, \chi_2 = 0$.  Additionally, one panel shows the fit for
the three EOSs and the extrapolation for $t/\tau_{_{\rm TOV}} \to 1$, as
discussed by \citep{Koeppel2019}.

\noindent\textit{Averaging method.~} As from its name, the
\textit{averaging} method simply averages the masses of the two binaries
that are closest to the critical threshold mass, with $M_{\rm sup}$ being
the \textit{supercritical} value and $M_{\rm sub}$ the
\textit{subcritical} one, \ie
\begin{align}
  \Mth := \frac{1}{2}\left(M_{\rm sup} + M_{\rm
    sub}\right)\,. \label{Aeq:Mthavg}
\end{align}
To distinguish between the two cases, we define as supercritical any
binary whose evolution of the normalised lapse sharply decreases to the
lower limit $\hat{\alpha}=0.1$ without a characteristic post-merger local maximum. On
the other hand, we define as subcritical any binary whose evolution of
the normalised lapse shows a local maximum after merger indicating,
therefore, that the merged object has contracted but also expanded at
least in one complete oscillation.

\begin{deluxetable*}{l|ccccc}[t!]
\setlength{\tabcolsep}{1.0mm}
  \tablecaption{Values of the fitting coefficients and their uncertainty
    in the functional ansatz (3) in the main text. The first three rows
    refer to the specific EOSs considered here and the last one to the
    quasi-universal expression. \label{tab:fits}}
  \tablehead{FIT & $a_1$ & $a_2$ & $a_3$ & $a_4$ & $a_5$ }
  \startdata
  \texttt{BHB}$\Lambda\Phi$ & $0.982 \pm 0.005$ & $~0.08 \pm 0.06$ & $0.14\pm0.02$ & $0.04\pm0.06$ & $-0.4\pm0.1$ \\
  \texttt{DD2}              & $0.997 \pm 0.007$ & $~0.13 \pm 0.08$ & $0.12\pm0.02$ & $0.04\pm0.08$ & $-0.3\pm0.1$ \\
  \texttt{TNTYST}           & $1.024 \pm 0.008$ & $~0.14 \pm 0.08$ & $0.12\pm0.03$ & $0.04\pm0.09$ & $-0.3\pm0.2$ \\
  \texttt{Univ}             & $1$               & $~0.11 \pm 0.07$ & $0.12\pm0.03$ & $0.07\pm0.09$ & $-0.3\pm0.2$
  \enddata
\end{deluxetable*}

Obviously, the averaging method is far simpler (it does not require any
fitting or extrapolation) but also does not provide any measure of how
close $M_{\rm sup}$ and $M_{\rm sub}$ are from $\Mth$. Interestingly, we
have measured $\Mth$ using the averaging method against the free-fall
method and obtained the same results with an average (maximum) difference
relative to the value obtained using the \textit{free fall} method of $<
1\%$ ($\sim 2 \%$). The situations in which the averaging method is to be
preferred is when the difference between $M_{\rm sup}$ and $M_{\rm sub}$
is $<1 \%$, in which case the error in measuring $\Mth$ would be
dominated by the evolution resolution. This was particularly useful in
the case of the \texttt{TNTYST} EOS, where the separation between the
supercritical and subcritical solutions was very small in some specific
mass ratios and spins.

Both of the methods outlined require a definition on whether or a not
a dataset is \textit{supercritical} or \textit{subcritical}.  In this work
we use the definition such that $\dot{\hat{\alpha}} |_{t_{\rm merg}}^{t_{\rm BH}}$
diverges directly. We note the use of the time derivative of $\hat{\alpha}$
is important especially in the highly asymmetric binaries as the
collapse behavior is less apparent when only considering $\hat{\alpha}$.
\end{document}